\documentclass[aps,prl,twocolumn,showpacs,preprintnumbers,amsmath,amssymb,superscriptaddress]{revtex4-1}
\usepackage{graphicx}
\usepackage{dcolumn}
\usepackage{bm}
\usepackage{color}
\usepackage{hyperref}
\newcommand{\nc}{\newcommand}
\nc{\be}{\begin{equation}}
\nc{\ee}{\end{equation}}
\nc{\bea}{\begin{eqnarray}}
\nc{\eea}{\end{eqnarray}}
\nc{\bean}{\begin{eqnarray*}}
\nc{\eean}{\end{eqnarray*}}
\nc{\mb}{\mbox}
\nc{\rnc}{\renewcommand}
\nc{\vk}{\mb{\bmk}}
\nc{\vp}{\mb{\bmp}}
\nc{\vn}{\mb{\bmn}}
\nc{\vq}{\mb{\bmq}}
\nc{\rr}{\mb{\bmr}}
\nc{\vz}{\hat {\mb{\bmz}}}
\nc{\vj}{\mb{\boldmath$j$}}
\nc{\vg}{\mb{\boldmath$g$}}
\nc{\x}{\mb{\boldmath$x$}}
\nc{\A}{\mb{\boldmath$A$}}
\nc{\va}{\mb{\boldmath$a$}}
\nc{\vs}{\mb{\boldmath$\sigma$}}
\nc{\vpi}{\mb{\boldmath$\pi$}}
\nc{\nab}{\nabla}
\nc{\X}{\sf x}
\usepackage{txfonts}
\usepackage{pxfonts}
\usepackage{graphicx,bm,units,yfonts}

\newcommand{\IM}{{\mathbb I}}

\newcommand{\ZM}{{\mathbb Z}}

\newcommand{\Cc}{{\mathcal C}}

\newcommand{\Kk}{{\mathcal K}}

\newcommand{\Ll}{{\mathcal L}}

\begin{document}

\title{Topological braiding of non-Abelian mid-gap defects in classical meta-materials}

\author{Yafis Barlas} 
\affiliation{Department of Physics, Yeshiva University, New York, NY 10016, USA}
\affiliation{Department of Physics, University of Nevada Reno, Reno, NV, 89522, USA}
\author{Emil Prodan}
\affiliation{Department of Physics, Yeshiva University, New York, NY 10016, USA}

\begin{abstract}
Non-trivial braid-group representations appear as non-Abelian quantum statistics of emergent Majorana zero modes in one and two-dimensional topological superconductors. Here, we  generate such representations with topologically protected domain-wall modes in a classical analogue of the Kitaev superconducting chain, with a particle-hole like symmetry and a $\ZM_2$ topological invariant. The mid-gap modes are found to exhibit distinct fusion channels and rich non-Abelian braiding properties, which are investigated using a T-junction setup. We employ the adiabatic theorem to explicitly calculate the braiding matrices for one and two pairs of these mid-gap topological defects. 
\end{abstract}

\maketitle

An important characteristic of a topological insulator is the emergence of boundary electron states, whenever a sample is halved. Similar topological effects can be engineered in classical meta-materials, where spectral gaps at finite frequencies support wave-guiding modes along domains or boundaries. This topological behavior has been demonstrated with photonic~\cite{Wang2009,Hafezi2013,PhysRevLett.114.223901}, phononic~\cite{PhysRevLett.103.248101,Mousavi2015,PhysRevB.97.054307} and mechanical~\cite{Kane2013,Nash14495,Susstrunk47,Paulose2015,Prodan2017} platforms. The electronic topological phases in condensed matter systems are classified by generic symmetries~\cite{SRFL2008,QiPRB2008,Kit2009,RSFL2010} and this classification also includes topological superconductors supporting Majorana quasi-particles as boundary modes \cite{kitaev_unpaired_2001,fu_superconducting_2008}. These boundary modes are far more exotic as they display non-Abelian braiding and statistics~\cite{kitaev_unpaired_2001,alicea_non-abelian_2011,mourik_signatures_2012}. While several classes of topological insulating systems have been implemented in meta-materials, the analogue of topological superconducting systems has not been yet realized. As such, an important question remained open, namely, whether topologically degenerate resonant levels can be engineered in meta-materials and if these degenerate modes can be manipulated by adiabatic deformations which ultimately result in non-trivial representations of the braid group.

 
In this Letter, we answer this question in the affirmative by designing a one-dimensional (1D) meta-material that exhibits non-trivial representations of the braid group.  Regardless of the quantum or classical setting, the critical ingredients for these representations are the adiabatic theorem, a non-trivial configuration space and a degenerate manifold of periodic solutions. We create them using the algorithmic map~\cite{PhysRevB.98.094310}, which translates any strong topological condensed matter system along with its corresponding symmetries, to an absolutely equivalent topological meta-material, built exclusively with passive components (such as magnetically coupled spinners~\cite{PhysRevMaterials.2.124203}). The result is a classical analogue of the particle-hole (PH) symmetric Kitaev chain~\cite{kitaev_unpaired_2001}, where topological mid-gap resonant modes can be stabilized by domain walls (DW) interpolating between trivial and topological regions. Since these modes are pinned at a fixed frequency and can be controlled by adiabatic displacements of DWs, one has the unique opportunity to explore the braiding of these point topological defects. 

 Our proposal is aided by two purely classical features with no quantum analogs, namely, the possibility to create two distinct DW configurations and the ability to control the ``superconducting order parameter.'' As a result, in our model, the number of different domain wall configurations resulting in $N$ mid-gap modes equals 
$2^{N/2}$ (see ~\cite{Supplemental} for the counting argument). Additionally, we discovered two distinct fusion channels, with the same fusion rules as Ising anyons. Aimed with these similarities, we describe how a braiding cycle can be implemented with a T-junction geometry, similar to Ref~\onlinecite{alicea_non-abelian_2011}. We show that, even though the equations of motion are quadratic in time, the braiding matrices can be computed using the adiabatic theorem~\cite{PhysRevB.80.115121}. This allows us to demonstrate, via explicit numerical calculations, that distinct exchange matrices result for different fusion channels. Using analytic arguments, based on the PH-symmetry of the system and localization of the mid-gap modes, we demonstrate that the braids are independent on the details of the implementation, hence topological, as long as the DWs are kept sufficiently far apart. Furthermore, we construct unitary representations of the braid group~\cite{Supplemental}, consistent with our numerical results. 

\begin{figure}[t]
\includegraphics[width=1.0\linewidth]{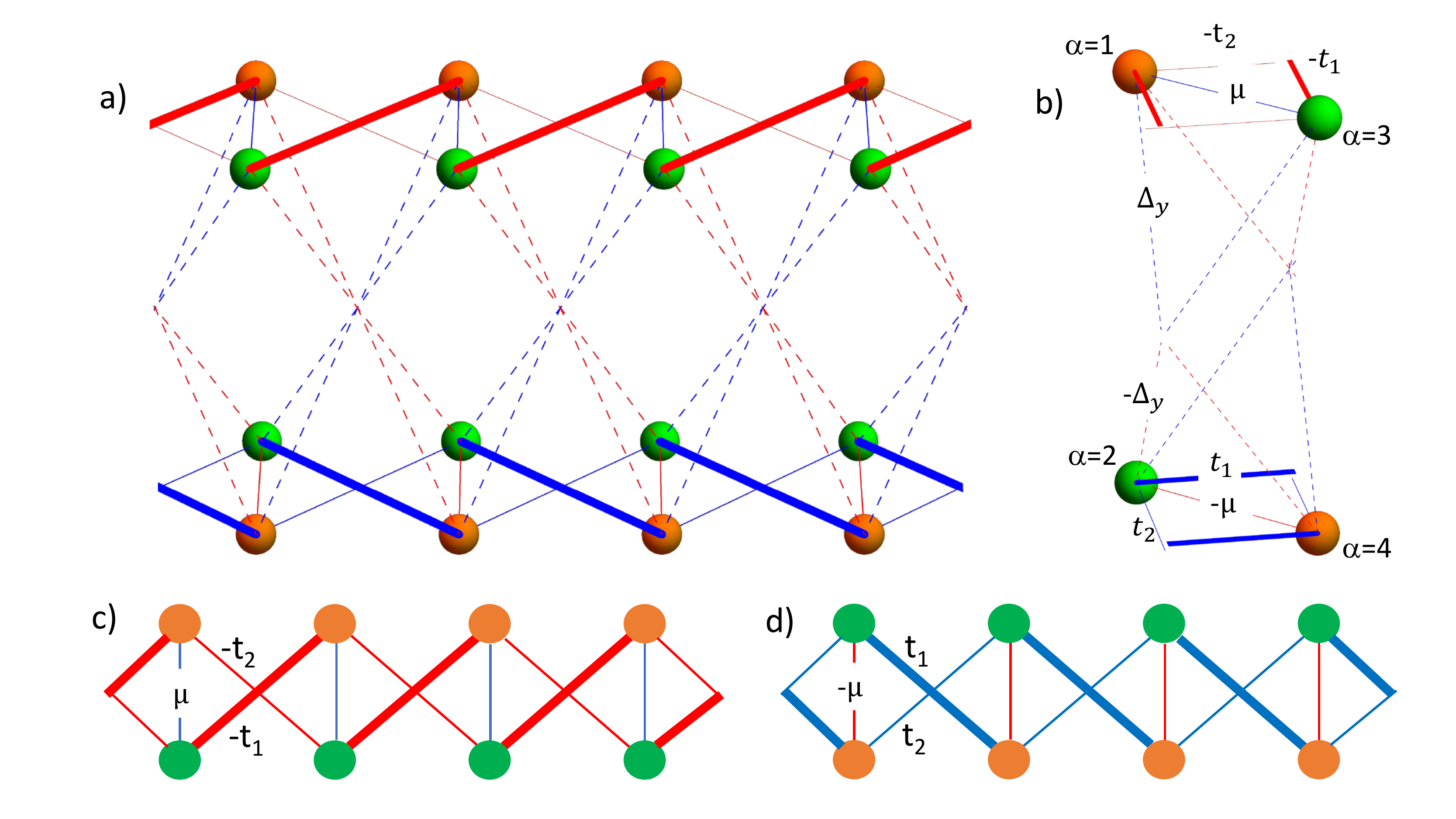}
\caption{ a) One-dimensional lattice with four identical resonators per unit cell. The resonators inside the $n$-th cell are labeled by $(n,\alpha)$, with $\alpha= \overline{1,4}$. The connections between the resonators represent the real valued couplings $d_{n m}^{\alpha,\beta}$ in Eq.~\ref{Eq:Couplings}, with $t_{1} = t +\Delta_{x}$ and $t_{2} =t-\Delta_{x}$. The positive (negative) signs of the couplings are color coded in blue (red). b) Detailed hopping pattern in the unit cell, viewed from a different angle. Figures c) and d) show the top and bottom layers respectively, which are coupled SSH ladders (see text for more details).}
\label{Fig:latticehopping}
\end{figure}

Before we proceed, let us emphasize that in 1D topological superconductors (D-class systems), Majorana zero-mode braiding requires adiabatic deformations that break all accidental symmetries, particularly, the time-reversal symmetry \cite{Supplemental}. As such, this task cannot be accomplished within the BDI-class, for which mechanical analogs already exist~\cite{Prodan2017}. Furthermore, in 1D systems, the $\ZM_2$ (as opposed to $\ZM$) stability is essential for implementing exchange for a sequence of four Majorana zero-modes~\cite{Supplemental}, hence this task cannot be accomplished in the AIII class either. Note that exchanging just two Majorana zero-modes presents no challenge and it has been implemented  with 1D classical chiral symmetric wires (class AIII) \cite{BorossArxiv2019}. As such, the $\ZM_2$ PH-symmetric D-class setting is not a convenient choice but rather a necessity, and this is the challenge for meta-material implementation which is addressed here. In 2-dimensions or in synthetic dimensions, non-Abelian geometric phases can be generated with other topological classes \cite{IadecolaPRL2016,SimenonovPRA2017}

The classical analogue of a fully general PH-symmetric Kitaev chain is shown in Fig.~\ref{Fig:latticehopping}. If we encode the degrees of freedom in the vector ${\bf Q} = \{q^{\alpha}_{n} \}$, the collective motion is determined by the quadratic Lagrangian:
\begin{equation}\label{Eq:Couplings} 
\Ll =\tfrac{1}{2}(\dot{\bm Q}^\dagger \dot{\bm Q} + \bm Q^\dagger D \bm Q), \ \ \bm Q^\dagger D \bm Q = \sum_{n,\alpha;m,\beta} d_{n m}^{\alpha \beta}\, q_n^\alpha \, q_m^\beta,
\end{equation} 
written in appropriate units, and the equation of motion becomes $\ddot{\bf Q} = - D {\bf Q}$. With the labelings and couplings from Fig.~\ref{Fig:latticehopping}, which were supplied by the map \cite{PhysRevB.98.094310}:
\begin{equation}
\label{Eq:Ham}
D = H + \omega_0^2 \, \IM, \quad  H = -\tau_{2} \otimes K,
\end{equation}
where $\omega_0$ is the pulsation of the un-coupled resonators, $\tau_2$'s is the off-diagonal Pauli matrix which couples the collective indices $ 
(1,2)$ with $ (3,4)$ (see Fig.~\ref{Fig:latticehopping} for details), and $K$ is the Kitaev Hamiltonian \cite{kitaev_unpaired_2001}:
\begin{equation}
\label{Eq:HamKitaev}
 K = \tfrac{\imath}{2}( \Delta_{x} \sigma_1 + \Delta_y \sigma_3)  \otimes (S - S^\dagger) - \hat{\sigma}_2 \otimes \big (\mu - \tfrac{t}{2}(S + S^\dagger) \big ),
\end{equation}
with complex order parameter $\Delta=\Delta_x + \imath \Delta_y$, on-site chemical potential $\mu$ and hopping parameter $t$ (fixed at 1). The shift operator of the lattice acts as $S\{q_n^\alpha\} =\{q_{n+1}^\alpha\}$ and $\sigma_i$'s are the Pauli matrices which separately act on indices $(1,2)$ and $(3,4)$. The PH-symmetry of \eqref{Eq:HamKitaev} is implemented by the complex conjugation $\Cc$, which remains a symmetry even if the parameters $\Delta$, $\mu$ are given site dependencies, {\it e.g.} to create domain walls.

\begin{figure}
\includegraphics[width=1.0\linewidth]{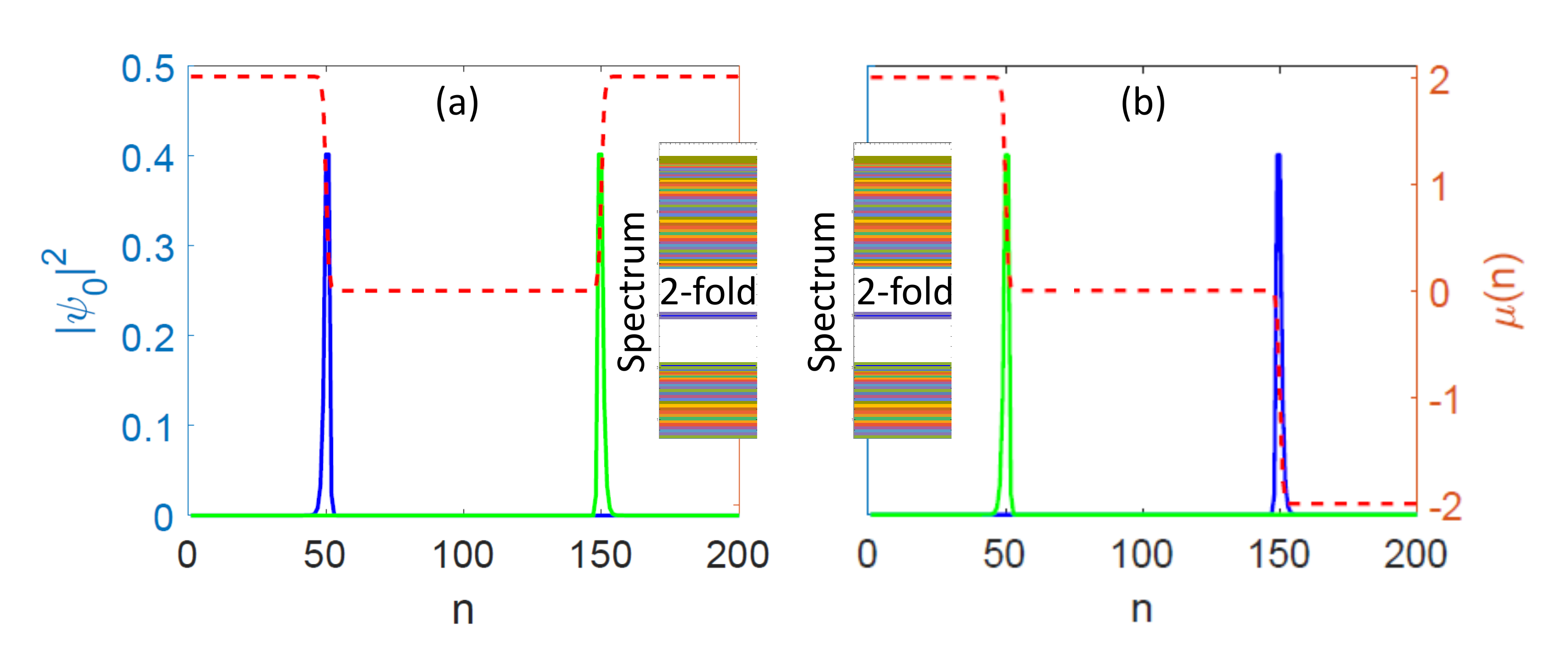}
\caption{\small (a) The $(++)$ domain wall configuration, generated by a spatial variation of $\mu$ (red-dashed line), interpolating between trivial ($\mu = 2$) and topological ($\mu =0$) phases. The phase interfaces trap two mid-gap modes, whose amplitudes are shown in blue and green. The inset displays the spectrum of $H_{+}$ calculated with this domain wall configuration. (b) Same as (a) but for the $(+-)$ domain wall configuration.}
\label{Fig:ZeroModes}
\end{figure}

All entries in $D$ are real valued, as it should be for passive meta-materials, but this comes at the expense of doubling the Kitaev model. Nevertheless, note the intrinsic symmetry $[H,S_{y}] = 0$, with $S_{y} = -i\hat{\sigma}_{2} \otimes \mathbb{I}$ ($S_y^2 = - \IM$), which fully decouples the two copies. Indeed, if: 
\begin{equation}
\Pi_\pm=\tfrac{1}{2}(\IM \mp \imath S_{y})= \pi_\pm \otimes \IM, \quad \pi_\pm = \tfrac{1}{2}\begin{pmatrix} 1 & \mp \imath \\ \pm \imath & 1 \end{pmatrix},
\end{equation} 
are the projections onto the symmetry sectors of $S_y$, then: 
\begin{equation}
H=H_+ \oplus H_- , \quad H_\pm = \Pi_\pm H \Pi_\pm = \pi_\pm \otimes K.
\end{equation} 
Furthermore, each reduced Hamiltonian $H_{\pm}$ obeys PH-symmetry $\Theta_{\rm PH} \, H_{\pm} \, \Theta_{\rm PH}^{-1} = -H_{\pm}$, with $\Theta_{\rm PH} = (\tau_1 \otimes \IM) \Cc$. The $\Pi_\pm$ sectors remain invariant under the dynamics, hence the mechanical system can be driven exclusively in one sector or the other \cite{Note1}. As such, from now on, we concentrate exclusively on $D_+ = H_+ + \omega_0^2 \, \Pi_+$, which apart from a shift, is unitarily equivalent with $K$ from \eqref{Eq:HamKitaev}.  Note that the time-reversal operation maps one sector into the other, hence $D_+$ is not constrained by this symmetry and this is why we can implement the T-junction cycle.

 The mapping procedure of Ref. \onlinecite{PhysRevB.98.094310} provides a fairly complex hopping pattern, which would be hard to guess at the onset. The effective PH-symmetry acts on the internal degrees of freedom of the lattice and, to gain more intuition, we start by separating the top and bottom layers by setting $\Delta_{y} =0$. The resulting top and bottom chains, with alternating $t_1$ and $t_2$ hopping patterns (see the Fig. 1 c $\&$ d), are classical analogues of coupled SSH models, with pairs of green/orange resonators supplying effective spin degrees of freedom. The coupling  $\mu$ provides an effective spin-orbit coupling, hence, in the limit of zero inter-layer coupling, our model is the classical analogue of a spin-orbit coupled wire. The inter-layer coupling $\Delta_y$ connects the same type of sites in opposite layers, and represents the classical analogue of the s-wave superconducting order parameter~\cite{footnote1}. Hence, Fig~\ref{Fig:latticehopping} represents the classical analogue of two Kitaev chains \cite{footnote11}, as argued above. The intrinsic symmetry of the lattice $[H,S_{y}] = 0$, acts on the top and bottom layers and represents a $\pi/2$-rotation in the internal effective spinor space. It fully decouples the two Kitaev chains.

The dispersion equation for a translational invariant configuration is $\omega_{\pm}^2 (k) = \omega_0^2 \pm \sqrt{(\mu -t \cos k )^2 + |\Delta|^2 \sin^2 k}$, hence the spectrum of $D_+$ is symmetric relative to $\omega_0^2$ and displays a gap as long as $\mu \neq |t|$. The system is in a topological (trivial) phase when $\mu < |t|$ ($\mu > |t|$).  The spectra of $D_+$ for two distinct DW-configurations,  created via spatial variation of $\mu$, are reported in Fig.~\ref{Fig:ZeroModes}. As expected, there are mid-gap resonant modes trapped at the interfaces between the topological and trivial phases, which are always present regardless of the particular monotonic profile of the DWs \cite{Note2}. Further spatial modulations of the parameter $\mu$ will result in more domain walls while maintaining the PH-symmetry at all times, hence one can nucleate an arbitrarily large number of spatially localized modes, whose pulsations are all pinned at $\omega_0$. Furthermore, by slowly changing the profile of $\mu$, we can coherently displace the topological modes in space while keeping the pulsation pinned at $\omega_0$. This supplies precise rules for how to modify the physical couplings and to ultimately implement this program in a laboratory \cite{Note3}, which is already unprecedented in a mechanical setting. 

\begin{figure}
\includegraphics[width=1.0\linewidth]{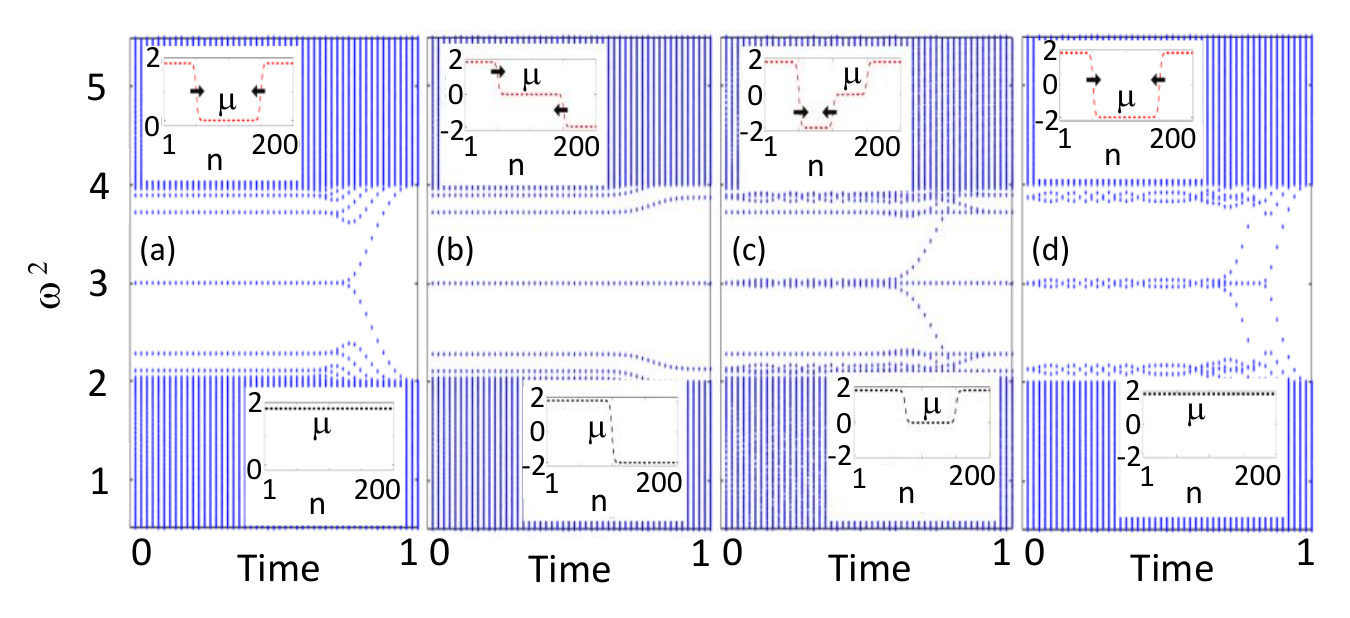}
\caption{\small The fusion results for different DW configurations, with insets indicating the configuration before and after the fusion: a) $\sigma \times \sigma =0$, b) $\sigma \times \sigma =\psi$, c) $\sigma \times \psi =\sigma$ and d) $\psi \times \psi =0$.}
\label{Fig:fusionrules}
\end{figure}

{\it Fusion rules.} For our classical system, there are two distinct ways of creating the DWs, by interpolating $\mu$ between $0$ and $\pm 2t$ as already exemplified in Fig.~\ref{Fig:ZeroModes}.  We think of these two possibilities as the manifestation of a non-trivial internal structure of same excitation $\sigma$ because both cases yield an identical energy spectrum. Furthermore, these two DW configurations can be both deformed into the topological-vacuum interface zero-mode without breaking the PH-symmetry or closing the bulk gap, e.g. by pushing the DWs to ends of the Kitaev chain.

In Fig.~\ref{Fig:fusionrules}(a), we report the spectral flow as two $\sigma$-excitations are adiabatically brought on top of each other via the $++$ channel. As one can see, the energies of the modes peel off from the mid-gap value and, when fused, the modes are completely lost to the bulk. Quite opposite, if the $\sigma$-excitations are fused in the $(+-)$ channel as in Fig.~\ref{Fig:fusionrules}(b), the modes persist and remain pinned in the middle of the gap. We have verified that this phenomena is not related to particular profiles of the domain-walls and found that the rule is robust as long as the DWs are sufficiently smooth. We think of the result of $+-$ fusion as a new type of excitation $\psi$. The remaining fusion rules are computed in Figs.~\ref{Fig:fusionrules}(c,d) and they match perfectly the ones for $SU(2)_2$ \cite{WangBook}: $\sigma \times \sigma = 0 + \psi$, $\sigma \times \psi = \sigma$, and  $\psi \times \psi =0$. These fusion rules are enabled by the ability to control the DW configurations of our classical system. 

\begin{figure}[t]
\includegraphics[width=\linewidth]{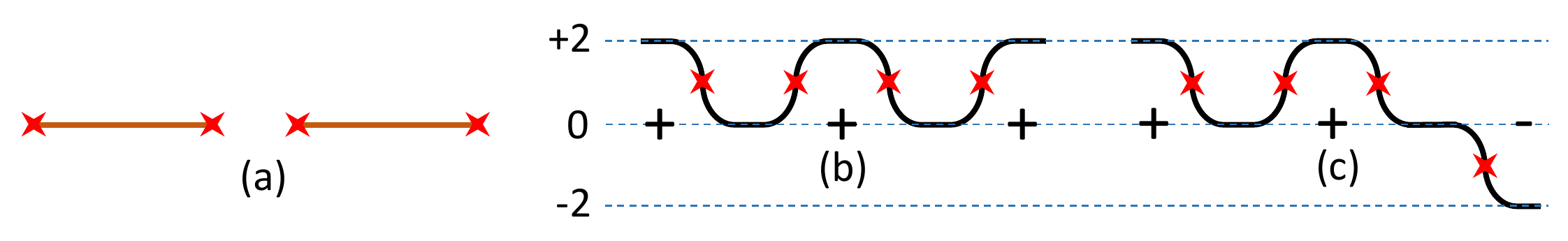}
\caption{\small (a) Two topological chains with sharp edges supporting mid-gap modes. (b/c) Smoothing of the interfaces results in different spatial profiles of $\mu$, labeled as $(+++)/(++-)$, and the system lands in the 0/$\psi$ sectors, respectively.}
\label{Fig:FusionSignificance}
\end{figure}

 To explain the significance of $0$, $\sigma$ and $\psi$ sectors and their fusion, we refer to Fig.~\ref{Fig:FusionSignificance}, which shows two topological chains with sharp edges supporting topological mid-gap modes. These modes can be braided only if the interface supporting them is smooth, otherwise the interface cannot be displaced in an adiabatic fashion (the Hamiltonian will display jumps no matter how slow the displacement!). The smoothing process can land the system in different DW-configurations, as shown in Fig.~\ref{Fig:FusionSignificance}(b,c). These configurations  always fall either in 0 or $\psi$ sectors (and $\sigma$ if the number of DWs is odd). To determine the sector, one can simply collapse all DWs on top of each other to find one of the three possible outcomes, $0$, $\sigma$ or $\psi$. Furthermore, the braidings can be implemented entirely in a single sector and we will give enough evidence that the braids depend in a non-trivial way on the DW-configurations. Lastly, when more strands like the ones in Fig.~\ref{Fig:FusionSignificance}(a) are added to the system, several distinct representations of the braid group can result, depending how the new strands are fused. The fusion rules supply a bookkeeping of these representations, in a manner which is yet to be determined.

{\it Braiding the DW-modes.}  In our classical mechanics setup, the equation of motion is quadratic in time, $-\partial_t^2 \bm Q = D(t) \bm Q$. Nevertheless, the adiabatic theorem \cite{Kato1950} still applies \cite{Supplemental}, more precisely, if the system is excited in any linear combination $\bm Q_{\omega_0}$ of mid-gap states, then at the end of the adiabatic cycle $\gamma(\tau)$ the system will oscillate as:
\begin{equation}
\bm Q(t) = {\rm Re}\big [ e^{\imath \omega_0 t} W_\gamma \bm Q_{\omega_0}\big ],
\end{equation}
where $W_\gamma$ depends entirely on path $\gamma$ inside the parameter space and can be conveniently computed as \cite{PhysRevB.80.115121}:
\begin{equation}\label{Eq:PMonodromy}
W_\gamma= \lim_{n \rightarrow \infty} P_{\gamma(\tau_n)}P_{\gamma(\tau_{n-1})} \ldots P_{\gamma(\tau_0)},
\end{equation}
where $\{\gamma(\tau_i)\}$ is a discretization of $\gamma$ and $P_{\gamma(\tau)}$ is the projection onto the mid-gap spectrum at the moment $\tau$.

\begin{figure}[t]
\includegraphics[width=\linewidth]{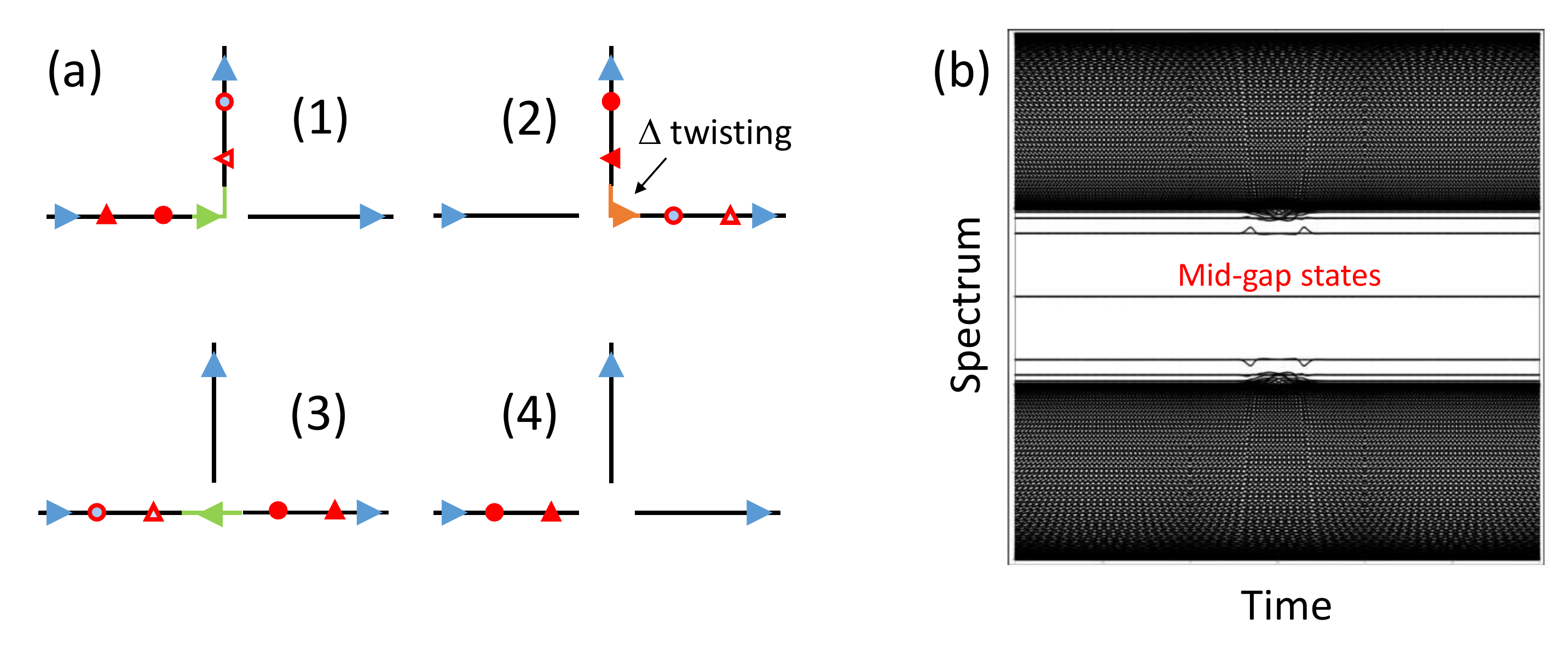}
\caption{\small (a). The T-junction braiding process, consisting of rigid slides of the DWs along the wires and adiabatic couplings/decouplings of the wires (shown as gree/orange segments). The closed (open) circles and triangles denote the initial (final) positions of the DWs after each step in the braiding process, while the arrows indicate the direction of the coupling. The coupling at step (2) requires a special twist of the $\Delta$ parameter, denoted by the orange connection. (b) Evolution of the spectrum during the full T-junction process, demonstrating that the mid-gap states remain spectrally separated at all times.}
\label{Fig:BraidingCycle}
\end{figure}

The braidings will be performed through different fusion channels, using all the available DW configurations and, as we shall see, this will result in different outcomes.  We first must make sure that our adiabatic braiding  cycles are closed, {\it i.e.} $H_+(t_{\rm init})=H_+(t_{\rm final})$ and that the bases we use are the same. For this, we always start/end our adiabatic cycles from a configuration where the zero modes are located at the clean ends of a chain and then we nucleate the desired smooth DW configurations. As for the adiabatic cycle, we employ the standard T-junction process illustrated in Fig.~\ref{Fig:BraidingCycle}(a) \cite{alicea_non-abelian_2011} but with one important difference. Note that $H_+$ is not symmetric under the inversion operation $\Kk$, $\Kk H_+(\Delta)\Kk^{-1}= H_+(-\Delta)$. Hence, the chains come with a definite orientation, shown by the blue arrow in Fig.~\ref{Fig:BraidingCycle}(a). If one insists on closing the braid cycle, then inherently two of the chains need to be fused in the wrong order, as it happens at step (2) in Fig.~\ref{Fig:BraidingCycle}(a). There, we are forced to connect two chains with opposite order parameters and, in order to keep the bulk-gap open, that connection requires a rotation in the complex plane of the order parameter $\Delta$. The precise expressions of the T-junction Hamiltonian is supplied in \cite{Supplemental}. Fig.~\ref{Fig:BraidingCycle}(b) reports the evolution of the spectrum of this adiabatic Hamiltonian during the whole T-junction cycle, demonstrating that the mid-gap states are spectrally separated at all times.

We now demonstrate the topological character of the braids and consider the case of two DWs. If the DWs are well separated, then the space of mid-gap modes accepts a very special basis $\{\Psi_1,\Psi_2\}$, with $\Psi_i$'s real valued and localized on one of the DWs. Independent of the location of the DWs in the T-junction geometry, such basis can be canonically generated. After an adiabatic exchange $1 \leftrightarrow 2$, the 1-dimensional spaces corresponding to $\Psi_{1,2}$ are swapped and, since the basis remains real at all times, the exchange matrix $U_{12}$ written in this basis takes an off diagonal form with real entries $\lambda_{12}$ and $\lambda_{21}$ and unitarity requires $\lambda_{12}^2 = \lambda_{21}^2 = 1$. This results in the possibilities $U_{12} = \pm \sigma_{1}$ or $\pm \imath \sigma_{2}$ hence, the monodromies are locked into one of these choices and continuous deformations cannot un-lock them, hence the topological character.

\begin{figure}
\begin{tabular}{cc}
\includegraphics[width=0.9\linewidth]{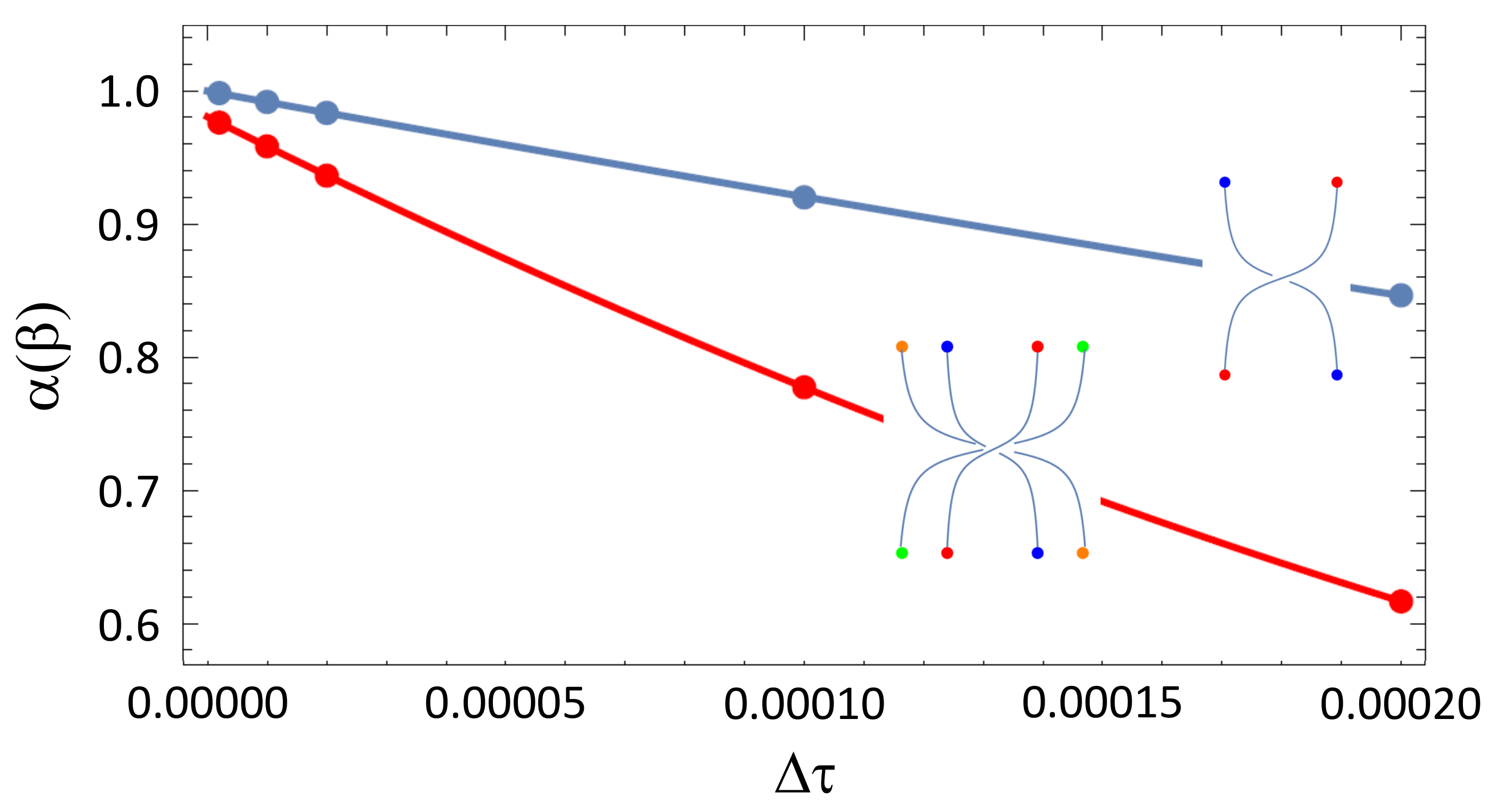}
\end{tabular}
\caption{\small Numerical evaluation of the braid cycles indicated in the diagram, as computed with \eqref{Eq:PMonodromy}. The graph reports $\alpha(\beta)$ (see text for definitions), which are plotted as function of the adiabatic type step. }
\label{Fig:NumericalBraids}
\end{figure}

The evaluation of the braid matrix was performed using the T-junction geometry and Eq~\ref{Eq:PMonodromy}, for the two braids shown in Fig.~\ref{Fig:NumericalBraids} via different DW configurations. The two mid-gap mode braid matrix via the $(++)$ channel can be expressed as $\alpha(\Delta t) [ \imath \sigma_{2} ]$, were $\alpha(\Delta \tau)$ plotted in Fig.~\ref{Fig:NumericalBraids} as a function of discretization step $\Delta \tau \to 0$. The fit to the scaling function, plotted in blue in Fig~\ref{Fig:NumericalBraids}, gives $\alpha (0) =0.99978 $, virtually converging to $1$. Via the $(--)$ channel, we find that $\alpha$ converges to $-1$. For the other more complicated braids in Fig~\ref{Fig:NumericalBraids}, we found the braid matrix to be given by $\beta(\Delta t)  [ \imath \sigma_2 \otimes \sigma_2 ]$ in the $(+++)$ channel and $ - \beta(\Delta \tau) [ \sigma_{1} \otimes \sigma_{1} ]$ in the $(+-+)$ channel, where $\beta(\Delta \tau) $ is plotted in red in Fig~\ref{Fig:NumericalBraids}, with $\beta(\Delta \tau \to 0) = 0.98054$. 

In a laboratory, for a pair of DWs, the two mid-gap modes can be loaded with arbitrary amplitudes and phases, leading to a collective oscillating state $\Psi(t) = {\rm Re} [\sum_{j=1,2}A_j\, \Psi_j \, e^{\imath (\omega_0 t + \phi_j)}]$. If one initiates two identical systems in the same load configuration, then one system can be braided while the other  can be left untouched. After the exchange, the resonators will oscillate as:
\begin{equation}
\Psi'(t)=\left \{
\begin{array}{l}
{\rm Re}( A_2 \Psi_2 \, e^{\imath\phi_2} - A_1 \Psi_1 \, e^{\imath \phi_1}  ) e^{\imath  \omega_0 t} \ \mbox{\small (++ ch.)} \\
{\rm Re}(A_1 \Psi_1 \, e^{\imath \phi_1}-A_2 \Psi_2 \, e^{\imath \phi_2} ) e^{\imath \omega_0 t} \ \mbox{(- - ch.)}
\end{array}
\right .
\end{equation}
which can be directly compared with $\Psi(t)$ using the second system. In particular, the emergence of the geometrical and topological phase of $\pi$ can be probed.

 In conclusion, we have supplied a classical analog of a fully general topological chain from class D, where all accidental symmetries are broken and the complex superconducting parameter $\Delta$ as well as the chemical potential $\mu$ can be manipulated by changing the coupling strengths between the resonators. This enabled us to create DWs supporting topological mid-gap states and to implement several braidings using T-junctions. We discovered that the braid matrices depend on the DWs configurations, which fall into three sectors displaying fusion rules similar to those for SU(2)$_2$. The braiding structure turned out to be far more interesting than anyone anticipated and this could be a proof that topological meta-materials could find applications in information processing.

\acknowledgments{Both authors acknowledge financial support from the W.M. Keck Foundation. Y.B. acknowledges additional support from UNR/VPRI start-up grant and E.P. from NSF grant DMR-1823800.}



\begin{thebibliography}{22}%

\bibitem{Wang2009} 
Z.~Wang, Y.~Chong, J.~D.Joannopoulos, M.~Soljacic, {\sl Observation of unidirectional backscattering-immune topological electromagnetic states}, Nature, {\bf 461}, 772 (2009).

\bibitem{Hafezi2013} 
M.~Hafezi, S.~Mittal, J.~Fan, A.~Migdall and J.~M. Taylor, {\sl Imaging topological edge states in silicon photonics} Nature Photonics, {\bf 7} 1001 (2013).

\bibitem{PhysRevLett.114.223901}
L.-H. Wu and X. Hu, {\sl Scheme for Achieving a Topological Photonic Crystal by Using Dielectric Material} Phys. Rev. Lett., {\bf 114} ,223901 (2015).

\bibitem{PhysRevLett.103.248101}
E.~Prodan, C.~Prodan, {\sl Topological Phonon Modes and Their Role in Dynamic Instability of Microtubules} Phys. Rev. Lett. {\bf 103},  248101(2009).

\bibitem{Mousavi2015}
S.~H. Mousavi, A.~B. Khanikaev and Z.~Wang, {\sl Topologically protected elastic waves in phononic metamaterials} Nature Communications {\bf 6}, 8682 (2015).

\bibitem{PhysRevB.97.054307}
R.~Chaunsali, C.-W. Chen and J.~ Yang, {\sl Subwavelength and directional control of flexural waves in zone-folding induced topological plates} Phys. Rev. B {\bf 97} 054307 (2018).

\bibitem{Kane2013}
C.~L. Kane and T.~C. Lubensky, {\sl Topological boundary modes in isostatic lattices} Nature Physics, {\bf 10}, 39 (2013).

\bibitem{Nash14495}
L.~M. Nash, D. Kleckner, A. Read, V. Vitelli, A.~M. Turner and W.~T.~M. Irvine,{\sl Topological mechanics of gyroscopic metamaterials}
  Proceedings of the National Academy of Sciences {\bf 112}, 14495 (2015).
	
\bibitem{Susstrunk47}%
R. S{\"u}sstrunk and S.~D. Huber, {\sl Observation of phononic helical edge states in a mechanical topological insulator}, Science, {\bf 349} 47 (2015).

\bibitem{Paulose2015}%
J. Paulose, B.~G.-g. Chen and V.~Vitelli,{\sl Topological modes bound to dislocations in mechanical metamaterials} Nature Physics {\bf 11}, 153, (2015).

\bibitem{Prodan2017}%
E.~Prodan, K.~Dobiszewski, A.~Kanwal, J.~Palmieri and C.~Prodan, {\sl Dynamical Majorana edge modes in a broad class of topological mechanical systems} Nature Communications, {\bf 8}, 14587 (2017).

\bibitem{SRFL2008} A.~P.~Schnyder, S.~Ryu, A.~Furusaki, A.~W.~W.~Ludwig, {\sl  Classification of topological insulators and superconductors in three  spatial dimensions}, Phys. Rev. {\bf B  78}, 195125 (2008).

\bibitem{QiPRB2008}  X.-L. Qi, T. L. Hughes, Shou-Cheng Zhang, {\sl Topological field theory of time-reversal invariant insulators}, Phys. Rev. B {\bf 78}, 195424 (2008).

\bibitem{Kit2009} A.~Kitaev, {\sl Periodic table for topological insulators and superconductors}, (Advances in Theoretical Physics: Landau Memorial Conference) AIP Conference Proceedings {\bf 1134}, 22-30 (2009).

\bibitem{RSFL2010}
S.~Ryu, A.~P. Schnyder, A.~Furusaki,  A.~W.~W. Ludwig, {\sl  Topological insulators and superconductors: tenfold way and  dimensional hierarchy}, New J. Phys. {\bf 12}, 065010 (2010).



\bibitem{kitaev_unpaired_2001}
A.~Y.  Kitaev, {\sl Unpaired Majorana fermions in quantum wires}, Physics-Uspekhi, {\bf 44}, 131 (2001).

\bibitem{fu_superconducting_2008}
L.~Fu and C.~L. Kane, {\sl Superconducting Proximity Effect and Majorana Fermions at the Surface of a Topological Insulator}, Physical Review Letters {\bf 100}, 096407 (2008).

\bibitem{alicea_non-abelian_2011}%
J.~Alicea, Y.~Oreg, G.~Refael, F.~v.Oppen and M.~P.~A. Fisher, {\sl Non-Abelian statistics and topological quantum information processing in 1D wire networks}, Nature Physics {\bf 7}, 412, (2011).

\bibitem{mourik_signatures_2012}%
V.~Mourik, K.~Zuo, S.~M. Frolov, S.~R. Plissard, E.~P. a.~M. Bakkers and L.~P. Kouwenhoven, {\sl Signatures of Majorana Fermions in Hybrid Superconductor-Semiconductor Nanowire Devices}, Science {\bf 336}, 1003, (2012).


\bibitem{PhysRevB.98.094310}%
Y.~Barlas and E.~ Prodan, {\sl Topological classification table implemented with classical passive metamaterials}, Phys. Rev. B, {\bf 98}, 094310 (2018).

\bibitem{PhysRevMaterials.2.124203}%
D.~J.Apigo, K. Qian, C. Prodan and E.~ Prodan, {\sl Topological edge modes by smart patterning}, Phys. Rev. Materials {\bf 2}, 124203 (2018).

\bibitem{Supplemental} See Supplemental Material at [URL]. 


\bibitem{PhysRevB.80.115121}
E.~Prodan and F.~D.~M. Haldane, {\sl Mapping the braiding properties of the Moore-Read state}, Phys. Rev. B, {\bf 80}, 115121, (2009).

\bibitem{BorossArxiv2019} 
P. Boross, J. K. Asboth, G. Szechenyi, L. Oroszlany, A. Paly, {\sl Poor man's topological quantum gate based on the Su-Schrieﬀer-Heeger model}, arXiv:1902.01358v1 (2019).

\bibitem{IadecolaPRL2016} T. Iadecola, T. Schuster, C. Chamon, {\sl Non-Abelian braiding of light}, Phys. Rev. Lett. {\bf 117}, 073901 (2016).

\bibitem{SimenonovPRA2017} L. S. Simeonov, N. V. Vitanov, {\sl Generation of non-Abelian geometric phases in degenerate atomic transitions},
Phys. Rev. A {\bf 96}, 032102 (2017).


\bibitem{Note1} A practical procedure for how to accomplish just that is supplied in~\cite{PhysRevB.98.094310}.

\bibitem{footnote1}  It is well established that the s-wave superconductivity in the presence of spin-orbit coupling, results in an effective p-wave superconductor.

\bibitem{footnote11}  With passive meta-materials, it is impossible to simulate a single Kitaev chain.

\bibitem{Note2} To implement smooth adiabatic processes, the DW profiles need to be rounded, {\it e.g.} as $\delta \mu\big(1+ \tanh\big (\tfrac{x-x_{\rm DW}}{\ell}\big)\big)$, where $x_{\rm DW}$ is the center of the wall, $\ell$ is the width of the wall and $\delta \mu$ is the full variation of $\mu$.

\bibitem{BraidBook} C. Kassel, V. Turaev, {\sl Braid Groups}, (Springer, Berlin, 2008).

\bibitem{Note3} Ref.~\onlinecite{PhysRevB.98.094310} explains in details how derive the physical couplings from a Hamiltonian written as in Eqs.~\eqref{Eq:Ham} and \eqref{Eq:HamKitaev}


\bibitem{WangBook} Z. Wang, {\sl Topological Quantum Computation}, (AMS, Providence, 2010).

\bibitem{Kato1950} T. Kato, {\sl On the adiabatic theorem of quantum mechanics}, J. Phys. Soc. J. Jpn. {\bf 5}, 435-439 (1950).
\end{thebibliography}

\end{document}